\documentclass[%
  reprint,        
  amsmath,amssymb,
  aps,            
  superscriptaddress,
  longbibliography
]{revtex4-2}

\usepackage[T1]{fontenc}
\usepackage[utf8]{inputenc}
\usepackage{graphicx}
\usepackage{xcolor}
\usepackage{hyperref}
\usepackage{siunitx}
\hypersetup{
  colorlinks=true,
  linkcolor=blue,
  citecolor=blue,
  urlcolor=blue
}

\usepackage{tikz}
\usetikzlibrary{arrows.meta,positioning,shapes.geometric,shapes.symbols}
\newcommand{\figref}[1]{Figure~\ref{#1}}

\newcommand{\agent}{\textcolor{black}{\textit{Accelerator Assistant}}}
\newcommand{\agentplain}{\textcolor{black}{Accelerator Assistant}}

\begin{document}

\title{Agentic Artificial Intelligence for Multistage Physics Experiments at a Large-Scale User Facility Particle Accelerator}

\author{Thorsten Hellert}\thanks{thellert@lbl.gov}
\author{Drew Bertwistle}
\author{Simon C.~Leemann}
\author{Antonin Sulc}
\author{Marco Venturini}

\affiliation{Lawrence Berkeley National Laboratory, Berkeley, California 94720, USA}

\date{January 16, 2026}

\begin{abstract}
We present the first language-model–driven agentic artificial intelligence (AI) system to autonomously execute multi-stage physics experiments on a production synchrotron light source.
Implemented at the Advanced Light Source particle accelerator, the system translates natural language user prompts into structured execution plans that combine archive data retrieval, control-system channel resolution, automated script generation, controlled machine interaction, and analysis. In a representative machine physics task, we show that preparation time was reduced by two orders of magnitude relative to manual scripting even for a system expert, while operator-standard safety constraints were strictly upheld. Core architectural features, plan-first orchestration, bounded tool access, and dynamic capability selection, enable transparent, auditable execution with fully reproducible artifacts. These results establish a blueprint for the safe integration of agentic AI into accelerator experiments and demanding machine physics studies, as well as routine operations, with direct portability across accelerators worldwide and, more broadly, to other large-scale scientific infrastructures.
\end{abstract}

\maketitle

\section{Introduction}
\label{sec:intro}

\begin{figure*}[ht!]
    \centering
    \includegraphics[width=1\linewidth]{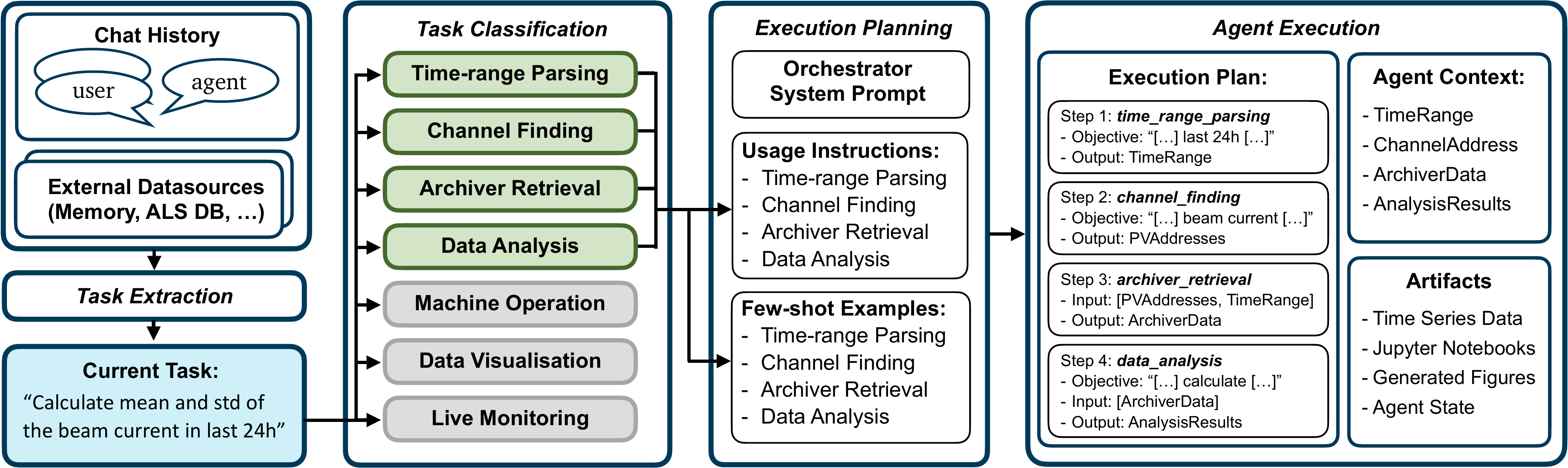}
    \caption{Overview of the agentic workflow. Multi-turn conversational input and external data sources are first processed into a structured task. Relevant capabilities are dynamically classified on each iteration of the interaction, and the description of the selected tools are passed to the execution planner. The planner generates a complete, inspectable execution plan with explicit dependencies, which is then carried out by the agent with context tracking and artifact management.}
    \label{fig:workflow}
\end{figure*}

Particle accelerators such as the Advanced Light Source (ALS)~\cite{Hellert2024zrj} are among the most complex scientific instruments, enabling frontier research in material science~\cite{Chen_2024,Tan_2025}, chemistry~\cite{Chandy_2025}, and biology~\cite{Ralston_2025}. Their operation requires continuous oversight by teams with expertise spanning accelerator physics~\cite{Wiedemann2015}, RF systems~\cite{Damerau2021}, magnets~\cite{Tanabe2005}, vacuum~\cite{MalyshevEtAl2019}, diagnostics~\cite{MintyZimmermann2003}, and controls~\cite{DiStefanoStubberudWilliams1997}. Because subsystem knowledge is distributed, operators frequently rely on domain specialists for troubleshooting, advanced tuning, or nonstandard experimental procedures. As user facilities, maximizing availability and protecting user beam are primary objectives; at the ALS, any beam interruption typically imposes a downtime of at least \SI{30}{\min}---and potentially several hours---with immediate consequences for dozens of concurrent experiments across more than 40 beamlines.

Many accelerator tasks extend beyond routine tuning, requiring custom scripts and deep subsystem knowledge. At the ALS, the control system exposes more than 230,000 process variables (PVs). Troubleshooting can be particularly demanding: unexpected faults lack predefined solutions, forcing operators to identify relevant channels, retrieve archive data, and assemble ad-hoc analysis under time pressure. Because each case is unique, preparation overhead and cognitive load are substantial, directly limiting machine availability and reducing scientific throughput across all beamlines. These challenges motivate the development of agentic systems that can translate user intent into structured, reproducible procedures while upholding strict facility safety constraints.

Recent advances in language models (LMs) point to a new class of agentic systems~\cite{russelnorvig2025ai} capable of bridging the gap between complex infrastructures and intuitive human interfaces. Beyond fluent text generation, LMs have been shown to support structured reasoning~\cite{wei2023chainofthought}, enabling them to decompose complex objectives into sequential steps. Building on this foundation, Toolformer~\cite{schick2023toolformer} demonstrated that models can be trained to call external tools, while ReAct~\cite{yao2023react} introduced reasoning–acting loops that tightly couple deliberation with execution. These ideas have since been extended to multi-agent orchestration~\cite{wu2023autogen}, memory-augmented systems~\cite{packer2023memgpt}, and graph-based planning frameworks~\cite{langgraph2025}, highlighting the potential of agentic AI to provide structured, inspectable execution. Yet most demonstrations remain confined to simulated or low-stakes domains; several domain-focused systems illustrate this diversity: ChemCrow augments LMs with chemistry-specific tools~\cite{bran2024chemcrow}, Co-scientist enables autonomous experimental planning in chemistry~\cite{boiko2023coscientist}, and CRISPR-GPT applies agentic orchestration to gene-editing workflows~\cite{qu2025crisprgpt}; and in synchrotron science, beamline prototypes such as VISION~\cite{Mathur_2025} have been explored, but reports of autonomous, multi-stage operation in production environments at large user facilities are still outstanding.

The constraints of accelerator facilities illustrate why: even small mistakes, such as a mistuned RF parameter or an incorrect magnet setting, can cause extended downtime, beam loss, or hardware damage, with immediate impact not only on machine health but also on dozens of concurrent experiments across all beamlines. These high-stakes conditions underscore the need for interfaces that are both intuitive and auditable.

Early explorations have already applied LMs to accelerator operation in targeted ways. GAIA~\cite{mayet2024gaia} introduced a prototype assistant at an R\&D linac, demonstrating how a LM could interface with logbooks, trigger control routines, and support operator workflows. Kaiser et al.~\cite{kaiser_LM} presented a proof-of-principle study where an LM performed beamline optics optimization from natural language prompts, directly comparing its performance against established optimization methods. Sulc et al.~\cite{Sulc2024xou} outlined a broader vision for integrating AI systems into accelerator control. While these works highlight the promise of LM–based interfaces, they remained limited in scope as either conceptual roadmaps or single-task demonstrations.

Here we advance beyond these prototypes by presenting the first deployment of a LM-based multi-agent system in a production synchrotron, the \agent. The system provides a natural language interface to an EPICS controls environment~\cite{EPICS} and extends beyond simple read/write access, enabling intuitive interaction with the control system through natural language: from a single user prompt it can retrieve archive data~\cite{archiver}, resolve PVs, generate and execute scripts, and analyze results. In a representative experiment (cf.~Section~\ref{section_from_query_to_exp}), the system autonomously prepares and executes a complete multi-stage procedure, reducing preparation time by two orders of magnitude relative to manual scripting by experts while strictly maintaining operator-standard safety constraints.

These capabilities are enabled by several architectural advances. Plan-first orchestration captures every task as an inspectable execution plan with optional operator approval. Dynamic capability filtering ensures stable scaling across a large tool inventory, while structured artifacts, including Jupyter Notebooks~\cite{Kluyver2016jupyter}, JSON outputs, and logs, provide reproducibility and transparency. Validating these methods under the strict availability and safety requirements of the ALS demonstrates that agentic AI can be used safely in high-stakes environments. The approach offers a blueprint for broader integration of LM–driven systems into large-scale scientific facilities, with direct portability to other synchrotrons, accelerators, and complex scientific infrastructures.

\section{ALS Agentic Control Framework}

The workflow of the agentic system follows a modular, capability-centric design that emphasizes separation of concerns following the Osprey framework~\cite{hellert2025osprey}. Each natural language fragment is first normalized into structured objects, ensuring that downstream components receive standardized inputs free of ambiguity. This modularity allows subsystems to be flexibly combined, enabling the framework to expand with new capabilities without requiring retraining or modification of existing components. An overview of the complete workflow is shown in \figref{fig:workflow}. All elements of the system, including source code, configuration files, and deployment scripts, are publicly available~\cite{als_assistant_repo}.

\begin{figure}[t]
    \centering
    \includegraphics[width=1\linewidth]{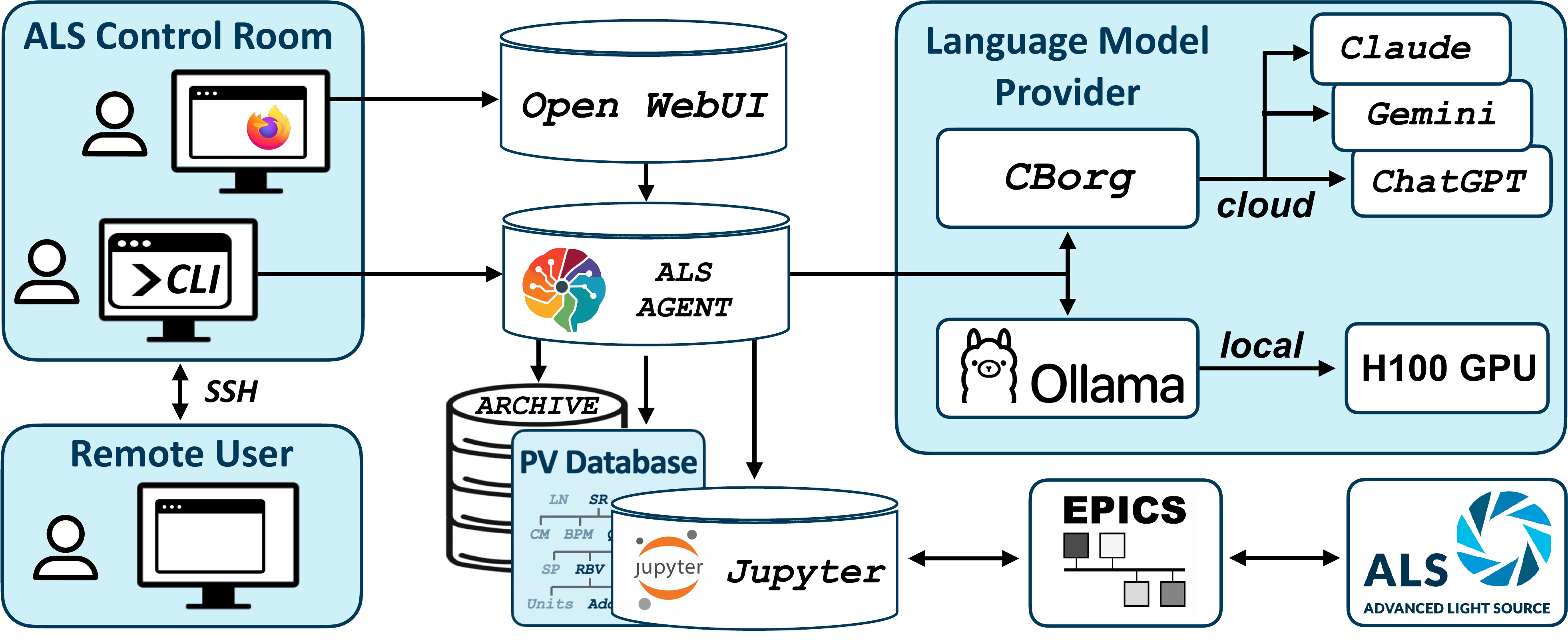}
    \caption{System architecture of the \agentplain. Control room and remote users access the system via a web interface (Open WebUI) or command line, which routes requests to the ALS Agent. The agent orchestrates connections to the PV database, archive data, and execution environments such as Jupyter. Model inference is performed either locally using Ollama or through cloud providers via the CBorg gateway. Integration with EPICS enables safe interaction with accelerator hardware at the ALS.}
    \label{fig:setup}
\end{figure}

Users interact with the system either through a command line interface or via the Open WebUI~\cite{open-webui}, accessible from every control room station as well as remotely over SSH. Authentication is tied to individual user identities, enabling the framework to maintain personalized context and memory across sessions. Multiple conversations can be managed in parallel, allowing users to organize distinct tasks or experiments into separate threads. As illustrated in \figref{fig:setup}, input is routed through the \agentplain, which orchestrates connections to the PV database, archive service, and Jupyter-based execution environments~\cite{jupyter_2021}.
Model inference is performed either locally using Ollama~\cite{ollama} on an H100 GPU node located within the control room network, or externally via the CBorg~\cite{cborg} gateway; a Lab-managed interface that routes requests to external providers such as ChatGPT~\cite{openai_chatgpt}, Claude~\cite{anthropic_claude}, or Gemini~\cite{google_gemini}. This hybrid architecture balances secure, low-latency on-premises inference with access to state-of-the-art foundation models, while integration with EPICS enforces operator-standard safety constraints for direct interaction with accelerator hardware.

At each turn, conversational input is translated into a concise and well-structured natural language task description that isolates objectives and removes redundancy. External knowledge sources, such as personalized memory stores tied to user identities, documentation, and accelerator databases, are incorporated to ground terminology and context. The resulting specification provides a clear objective, giving downstream components an unambiguous basis for execution while remaining human-readable.

The system organizes its functional units into modular \textit{capabilities}: self-contained tools for data retrieval, machine interaction, or analysis—that can be composed as needed for a given task. Capabilities are then classified for relevance to the current task. Each capability is evaluated independently, with the classification posed as a binary decision using few-shot, capability-specific examples. Only those deemed relevant are passed forward, preventing prompt inflation and decoupling task complexity from the size of the overall capability inventory, which ensures efficient operation and allows the framework to scale as new capabilities are added.

Execution proceeds via a plan-first orchestration strategy: before any tool is called, the system generates a complete execution plan that encodes explicit input–output dependencies. This separation of planning and execution ensures that logic remains transparent, serializable, and subject to inspection or modification. Plans also provide natural checkpoints where safety gates can be enforced: operators or automated validators may review inputs, outputs, and dependencies prior to the initiation of any action.

The execution environment is modular and containerized via Podman~\cite{podman}, ensuring reproducibility across both development workstations and production control room servers. Core components, including the agent, Open WebUI, and Jupyter services, run in isolated containers, allowing for consistent deployment, straightforward upgrades, and strict separation of privileges. Reliability features include checkpointing, structured error classification, and bounded retries with automatic re-planning when required. Human-in-the-loop interrupts are supported, allowing users to inspect and approve plans, code, or memory operations before side effects occur. Every run produces structured artifacts (including logs, JSON outputs, and Jupyter Notebooks) that provide a complete provenance trail, enabling reproducibility, auditing, and further development of workflows. In addition, the agent can materialize monitoring artifacts from natural language requests: for example, the prompt “Monitor the beam current and RF cavity temperature” generates a CS-Studio Phoebus Data Browser file~\cite{Shroff2023fzw}, a widely used control room toolkit with deep EPICS integration and the standard interface at the ALS, pre-configured to query the Archive Appliance for historical context while updating in real time. These auto-generated panels both eliminate PV lookup and data entry overhead and persist as reusable control room resources, yielding a practical speedup for routine observation and troubleshooting.

Although this work is demonstrated on the ALS, a mature and well-characterized storage ring, recent developments in the underlying Osprey framework extend the approach to a broader range of accelerator environments. The integration of advanced agentic code-generation modules, such as the Claude Code SDK~\cite{anthropic2025_claude_agent_sdk}, allows each facility to provide example code, conventions, and safety patterns that guide the Python generator to follow local operational practices. In parallel, Osprey’s safety layer supports configurable PV boundary limits as well as write blacklists and whitelists, enabling facilities to define safe operating ranges without modifying the framework. These configuration-level mechanisms make the system adaptable to variations in actuator behavior, diagnostic conditions, and metadata quality, providing a practical path for deployment at facilities with different levels of maturity.

\section{From Query to Experiment}
\label{section_from_query_to_exp}

To convey the capabilities of the \agentplain, we focus on a non-routine but practically important machine physics task. Such procedures are complex enough to require custom scripting but occur too infrequently for dedicated solutions to exist, making them an ideal proving ground for agentic control. In this case the task involves insertion devices (IDs)~\cite{Onuki2003er}: tunable undulator magnets whose gap settings strongly affect both machine optics and delivered photon beams.

As an illustration, the user made the following request:

\begin{quote}
\textit{``Get the minimum and maximum value of all ID gap values in the last three days. Then write a script which moves each ID from maximum to minimum gap and back while measuring the vertical beam size at beamline 3.1. Sample the gap range with 30 points, wait 5\,s after each new setpoint for the ID to settle and measure the beam size 5 times at 5\,Hz. Return a hysteresis plot beam size vs gap.''}
\end{quote}

This demonstration highlights both the scope and the impact of the system. While the beam-based measurement sequence itself necessarily requires about an hour, the preparation effort is reduced from what would typically take several hours of manual scripting and debugging to only a few minutes from a single natural language prompt---a speedup of two orders of magnitude. This efficiency results from the system’s ability to transform free-form user input into a structured execution plan that decomposes the request into modular steps.

The resulting execution plan organizes the task into a small number of safety-gated stages that can be inspected and, depending on configuration, approved either at every step or only for sensitive operations. In the present deployment at the ALS, the system has been configured to require operator approval for all write access to the control system. The stages comprise time-range normalization, PV resolution against the accelerator middle layer, archive retrieval, data analysis, controlled machine interaction, and visualization.

The workflow begins with time range parsing, a dedicated language processing task performed by a lightweight model to ensure low latency. Rather than relying on brittle pattern matching, the model flexibly interprets natural language fragments such as ``last three days'' and normalizes them into standardized \texttt{datetime} ranges that downstream services, such as archive queries, can directly process.

\begin{figure}[ht]
    \centering
    \includegraphics[width=1\linewidth]{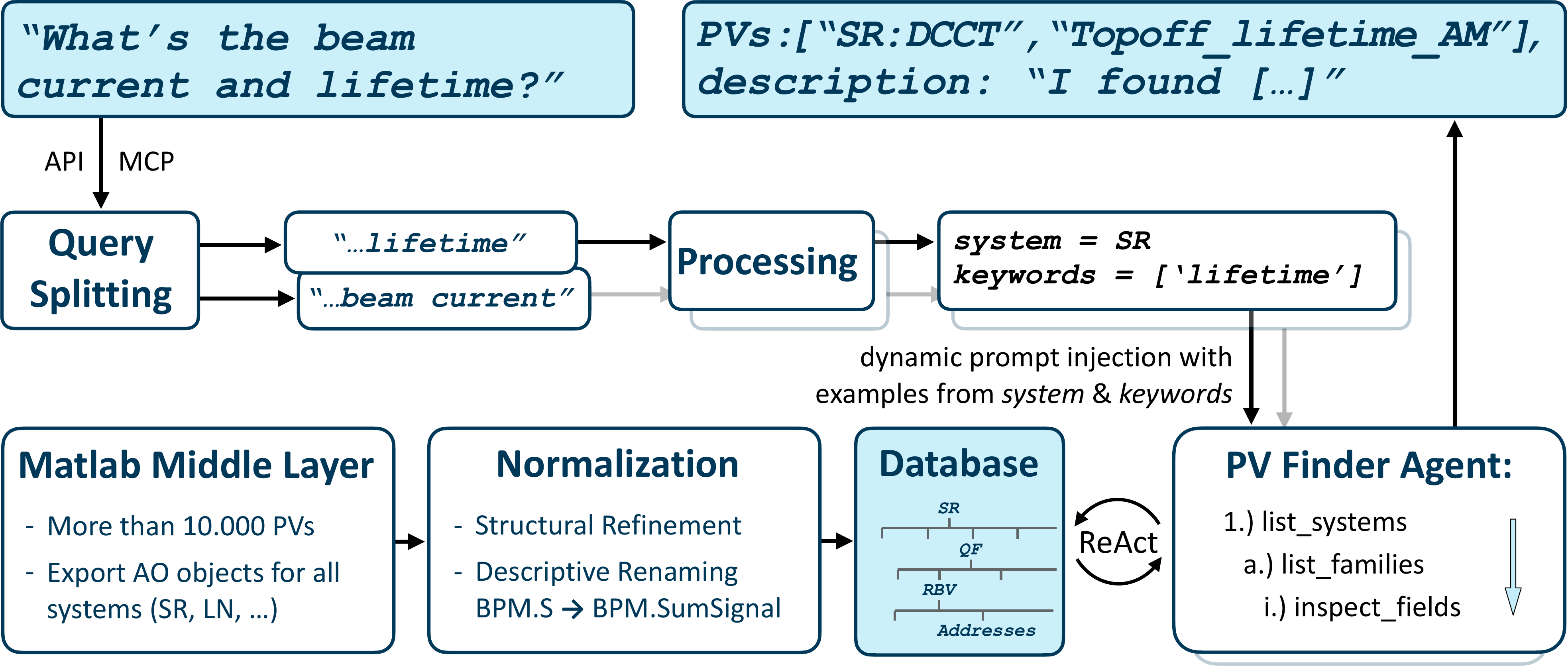}
    \caption{Workflow of the PV Finder subsystem. A normalized export of the Matlab Middle Layer Accelerator Object provides the data model, which the agent explores through a strictly bounded API. Natural language queries are split into atomic intents, preprocessed to extract systems and keywords, and resolved into specific EPICS PVs.}
    \label{fig:pv_finder}
\end{figure}

A central step is PV resolution, handled by the PV Finder subsystem (\figref{fig:pv_finder}). A normalized export of the MATLAB Middle Layer (MML)~\cite{Portmann_MML} Accelerator Object (AO) provides the underlying data model, including approximately 10,000 key PVs across all accelerator subsystems. Because the MML is implemented at most synchrotron light sources worldwide, this foundation makes the approach naturally transferable to other storage rings, with only minor refinements and descriptive renaming required to maintain a consistent, interpretable database structure for the language model. User queries are split into atomic intents, preprocessed to extract target systems and keywords, and then resolved into specific PVs by a ReAct-style agent restrained to a strictly bounded API. This tool-bounded exploration guarantees auditability while grounding ambiguous user terminology, such as ``ID gap'' or ``beam size'', into precise EPICS channel names.

Once PVs and time ranges are resolved, archive retrieval reduces to a straightforward API call to the archiver client. Input mapping is handled automatically by the orchestrator, returning time series data for all relevant IDs without user intervention.

\begin{figure}[h]
    \centering
    \includegraphics[width=1\linewidth]{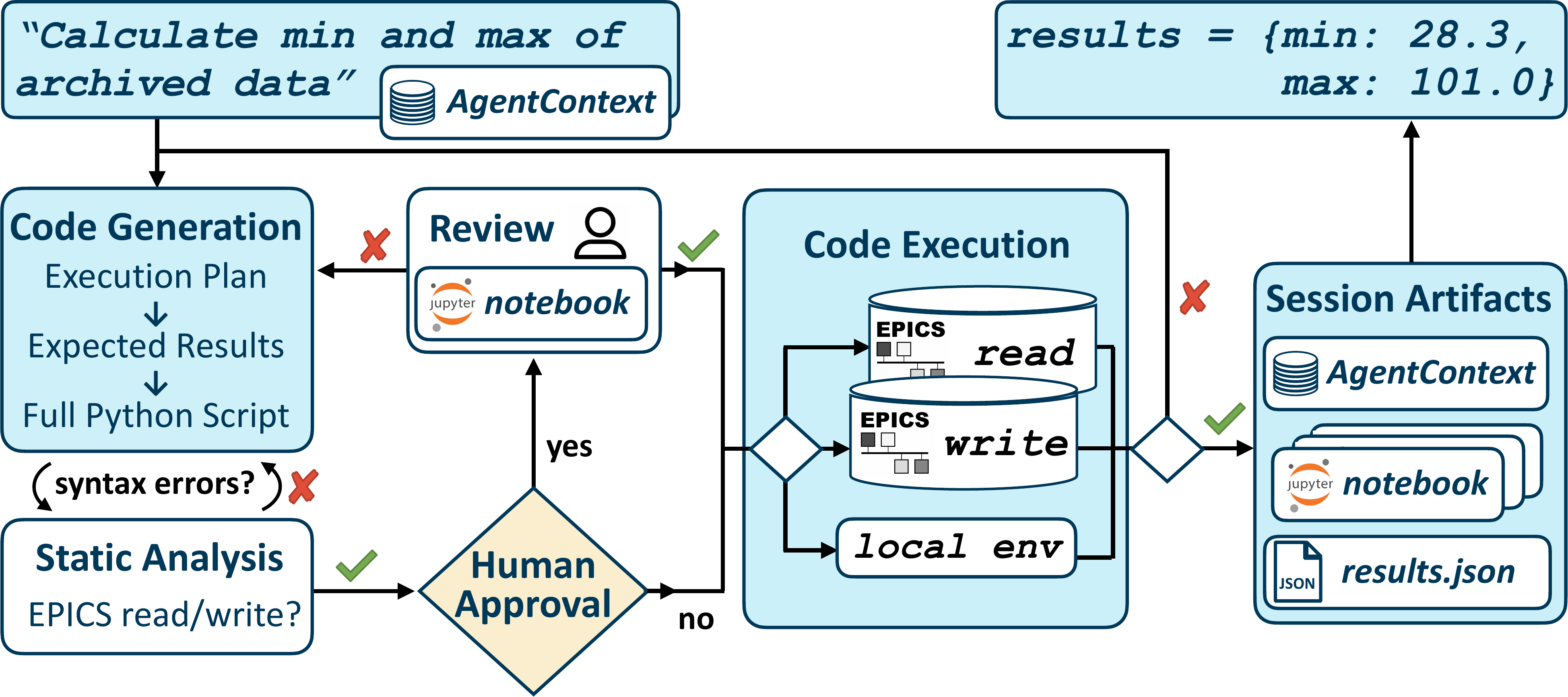}
    \caption{Pipeline for controlled Python code execution in the \agentplain. Natural language tasks are translated into a plan, results schema, and then Python code, which can dynamically access the agent context, is statically analyzed, and may be reviewed by a human operator. Execution is typically confined to containerized Jupyter kernels with strict read/write policies, and every run produces session artifacts (context, notebooks, JSON) for full reproducibility.}
    \label{fig:python_executor}
\end{figure}

Execution then proceeds through dynamically generated Python scripts (\figref{fig:python_executor}). To make this stage robust, code generation is decomposed into three successive model calls rather than a single direct request, which can be brittle and prone to over-design. First, the model produces a high-level plan of what the script should achieve. Second, this plan and the user’s objective are used to generate a structured JSON schema specifying the expected results. Finally, conditioned on both the plan and schema, the model produces the minimal Python code required to carry out the task. Scripts are executed inside containerized Jupyter kernels with strict read/write policies, supporting two modes: read-only (analysis and visualization) and write-enabled (machine interaction), the latter requiring operator approval by configuration (default policy), with read-only analysis as the baseline mode. All code may be reviewed prior to execution, and every run produces structured artifacts (Jupyter notebooks, JSON results, and figures) for provenance and reproducibility.

In the aforementioned example, the Python executor is invoked three times. First, it computes the minimum and maximum gap values from archived data. Second, it generates and runs a scan script that sweeps ID gaps between these values while recording synchronized beam size measurements. Third, it visualizes the acquired data in the form of a hysteresis plot, confirming the absence of significant beam size hysteresis. Importantly, these are not just three generic code-generation calls but invocations of distinct, specialized capabilities for data analysis, machine operation, and data visualization. Each capability follows the same modular architecture and execution flow illustrated in \figref{fig:python_executor}, yet is guided by tailored prompts that reflect its specific domain. This modularity has proven essential for achieving robust performance in the control room, ensuring that natural language user requests can be translated into complex, end-to-end experimental procedures with full auditability.

\begin{figure}[ht!]
\centering
\includegraphics[width=1\linewidth]{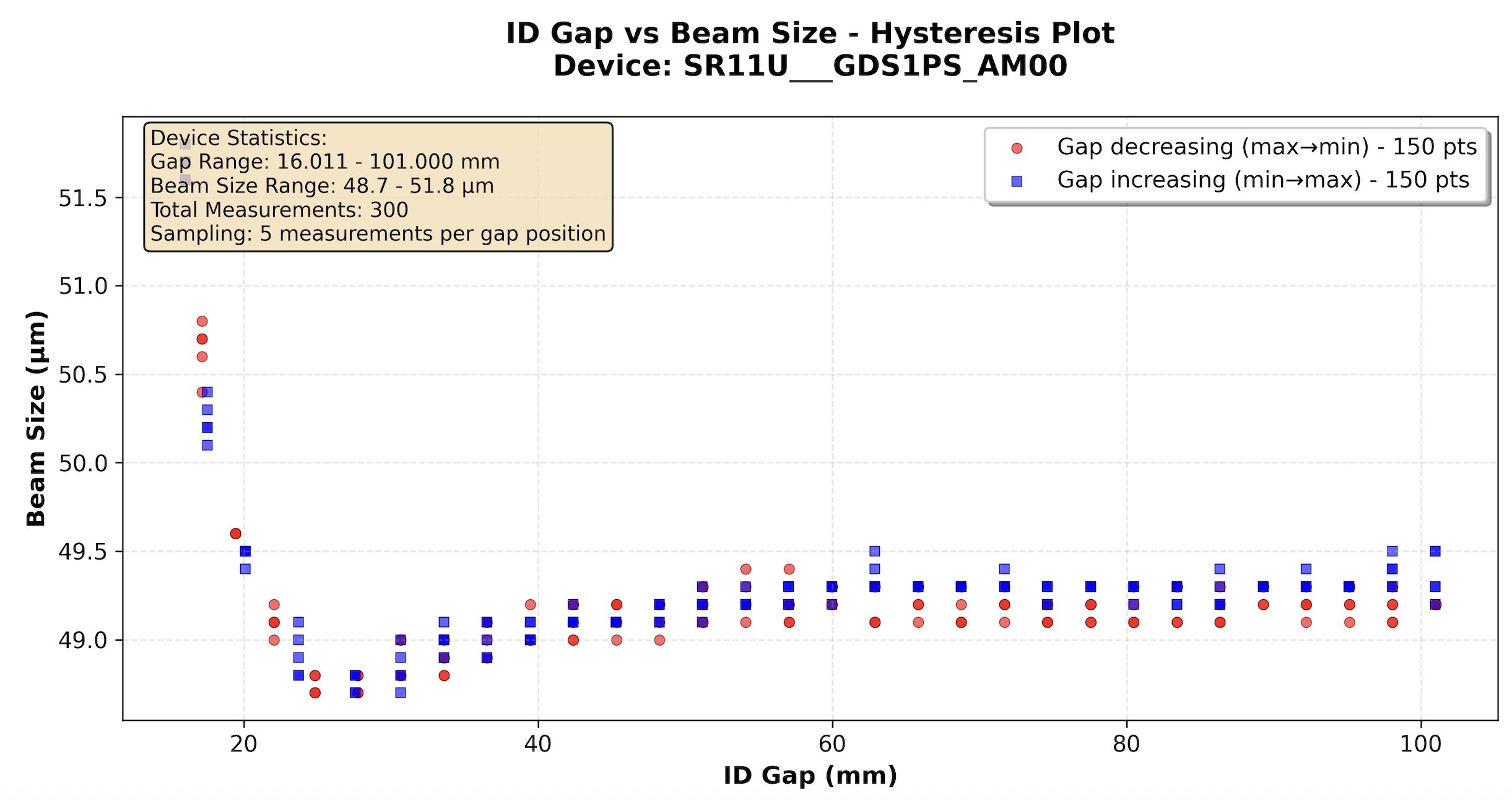}
\caption{Example output of the \agentplain: hysteresis measurement of ID gap versus vertical beam size at the ALS. The execution plan generated by the agent combined historical range extraction, automated script generation, and real-time machine control. The agent performed a 30-point bidirectional gap sweep with 5 repeated measurements per point, producing the plot shown here for one device. This figure illustrates the final output of the agentic workflow, where every step, from parsing natural language to data retrieval, machine control, and plotting, was generated and executed automatically.}
\label{fig:als_expert_hysteresis_plot}
\end{figure}

By chaining these modular capabilities, the system not only retrieves and analyzes archived data, but also orchestrates real-time machine operation. The result here is a series of consistent and publication-ready plots across all devices. An example is shown in \figref{fig:als_expert_hysteresis_plot}, demonstrating the expected absence of hysteresis in the vertical beam size.

This experiment illustrates how the \agentplain~converts natural language requests into fully executed physics measurements. Beyond this specific example, the same architecture applies to a wide range of machine physics tasks, providing a reproducible bridge from user intent to automated execution.

\section{Conclusion}
\label{sec:conclusion}
We have presented the first deployment of a LM–based agentic system executing a multi-stage physics experiment on a production synchrotron light source storage ring. By integrating directly with the EPICS control system and Archive Appliance, the \agentplain~demonstrates that natural language user requests can be converted into safe, auditable, and fully automated experimental procedures. The agent reduces preparation time by two orders of magnitude while preserving operator-standard safety constraints, and it produces reproducible artifacts that support inspection and trust.

Key architectural features, including plan-first orchestration, bounded tool access, and dynamic capability selection, ensure that the framework scales with growing functionality while remaining transparent and portable across accelerator facilities.

These results establish a blueprint for the safe integration of agentic AI into scientific operations. Beyond synchrotrons, the demonstrated principles are broadly applicable to other large-scale experimental facilities, where automation, transparency, and reproducibility are equally critical.

\section*{Acknowledgments}

This research leveraged the CBorg AI platform and resources provided by the IT Division at Lawrence Berkeley National Laboratory. We gratefully acknowledge Andrew Schmeder for his consistent responsiveness and support, which ensured that CBorg served as an invaluable resource for the development of this framework and NLP efforts in general at ALS.

We are grateful to Alex Hexemer, Hiroshi Nishimura, Fernando Sannibale, and Tom Scarvie (LBNL) for stimulating discussions and continued support, and to Frank Mayet (DESY) for sharing insights from his pioneering Gaia prototype, which guided the early trajectory of agentic AI at ALS.

We further acknowledge the use of AI tools during the preparation of this work. Cursor, primarily with Claude 4, was employed extensively during development.

This work was supported by the Director of the Office of Science of the U.S.~Department of Energy under Contract No. DE-AC02-05CH11231.

\bibliographystyle{apsrev4-2}
\bibliography{references}

\end{document}